\PassOptionsToPackage{dvipsnames}{xcolor}
\PassOptionsToPackage{hypertexnames=false}{hyperref}
\documentclass[aps,prl,reprint,superscriptaddress,nopreprintnumbers,nobibnotes,floatfix,longbibliography,footinbib]{revtex4-2}

\usepackage{newtxtext,newtxmath}
\usepackage{microtype}
\usepackage{graphicx,placeins}
\usepackage[normalem]{ulem}
\graphicspath{figures/}

\usepackage{orcidlink}

\usepackage{titlesec}
\titleformat{\section}[runin]{\itshape\addtocounter{section}{1}}{\thesection}{0pt}{}[---]
\titlespacing{\section}{\parindent}{0pt}{0pt}
\titleformat{\subsection}[runin]{\upshape}{\thesubsection}{0pt}{}[:]
\titlespacing{\subsection}{\parindent}{0pt}{1em}

\makeatletter

\renewcommand\tagform@[1]{\maketag@@@{\ignorespaces#1\unskip\@@italiccorr}}
\renewcommand\theequation{(\oldtheequation)}
\makeatother
\newcommand{\Illustris}{Illustris}
\newcommand{\TNG}{TNG}
\newcommand{\TNGfifty}{TNG50}
\newcommand{\Subfind}{\textsc{SubFind}}
\newcommand{\vLSR}{\ensuremath{v_\mathrm{\scriptscriptstyle LSR}}}
\newcommand{\vesc}{\ensuremath{v_\mathrm{esc}}}
\newcommand{\MLSR}{\ensuremath{M_\mathrm{LSR}}}
\newcommand{\xsec}{\ensuremath{\sigma_{\mathrm{\scriptscriptstyle SI}}}}
\newcommand{\kms}{\ensuremath{\text{km}~\mathrm{s}^{-1}}}
\newcommand{\GeV}{\ensuremath{\text{Ge\kern-0.25exV}}}
\newcommand{\GeVcm}{\ensuremath{\GeV~\mathrm{cm}^{-3}}}
\newcommand{\GeVc}{\ensuremath{\GeV~c^{-2}}}
\newcommand{\osim}{\mathord{\sim}}

\newcommand{\Msun}{\ensuremath{\mathrm{M}_\odot}}
\newcommand{\Rsun}{\ensuremath{R_{\kern-0.1ex\odot}}}
\newcommand{\Rm}{\ensuremath{R_{\kern-0.2ex M}}}
\newcommand{\Mdyn}{\ensuremath{M_\mathrm{dyn}}}
\newcommand{\Mstar}{\ensuremath{M_\star}}

\newcommand{\vb}[1]{\ensuremath{\mathbf{#1}}}
\newcommand{\tvb}[1]{\ensuremath{\tilde{\mathbf{#1}}}}
\definecolor{darkblue}{HTML}{2E3092}
\hypersetup{colorlinks=true,allcolors=darkblue}

\usepackage{fancyhdr}
\begin{document}

\title{Dark Matter Velocity Distributions for Direct Detection:\\Astrophysical Uncertainties Are Smaller Than They Appear}

\author{Dylan Folsom~\orcidlink{0000-0002-1544-1381}}
\email[Contact author: ]{dfolsom@princeton.edu}
\affiliation{Department of Physics, \href{https://ror.org/00hx57361}{Princeton University}, Princeton, NJ 08544, USA}

\author{Carlos Blanco~\orcidlink{0000-0001-8971-834X}}
\altaffiliation{NASA Einstein Fellow}
\affiliation{Institute for Gravitation and the Cosmos, \href{https://ror.org/04p491231}{The Pennsylvania State University}, University Park, PA 16802, USA}
\affiliation{Department of Physics, \href{https://ror.org/00hx57361}{Princeton University}, Princeton, NJ 08544, USA}
\affiliation{\href{https://ror.org/05f0yaq80}{Stockholm University} and The Oskar Klein Centre for Cosmoparticle Physics, Alba Nova, 10691 Stockholm, Sweden}

\author{Mariangela Lisanti~\orcidlink{0000-0002-8495-8659}}
\affiliation{Department of Physics, \href{https://ror.org/00hx57361}{Princeton University}, Princeton, NJ 08544, USA}
\affiliation{Center for Computational Astrophysics, \href{https://ror.org/00sekdz59}{Flatiron Institute}, New York, NY 10010, USA}

\author{Lina Necib~\orcidlink{0000-0003-2806-1414}}

\author{\\Mark Vogelsberger~\orcidlink{0000-0001-8593-7692}}
\affiliation{Department of Physics and Kavli Institute for Astrophysics and Space Research, \href{https://ror.org/042nb2s44}{Massachusetts Institute of Technology}, Cambridge, MA 02139, USA}

\author{Lars Hernquist~\orcidlink{0000-0001-6950-1629}}
\affiliation{\href{https://ror.org/03c3r2d17}{Center for Astrophysics, Harvard \& Smithsonian}, 60 Garden Street, Cambridge, MA 02138, USA\vspace{\baselineskip}}

\received{28~May~2025; accepted 27~October~2025; published 20~November~2025}

\begin{abstract} 
The sensitivity of direct detection experiments depends on the phase-space distribution of dark matter near the Sun, which can be modeled theoretically using cosmological hydrodynamical simulations of Milky Way--like galaxies. However, capturing the halo-to-halo variation in the local dark matter speeds---a necessary step for quantifying the astrophysical uncertainties that feed into experimental results---requires a sufficiently large sample of simulated galaxies, which has been a challenge. In this Letter, we quantify this variation with nearly 100~Milky Way--like galaxies from the \TNGfifty{} simulation, the largest sample to date at this resolution. Moreover, we introduce a novel phase-space scaling procedure that endows every system with a reference frame that accurately reproduces the local standard-of-rest speed of our Galaxy, providing a principled way of extrapolating the simulation results to real-world data. 
The ensemble of predicted speed distributions is well characterized by the standard halo model, a Maxwell--Boltzmann distribution truncated at the escape speed, though the individual distributions can deviate from it, especially at high speeds.
The dark matter--nucleon cross section limits placed by these speed distributions vary by $\osim60\%$ about the median. This places the 1$\sigma$ astrophysical uncertainty at or below the level of the systematic uncertainty of current ton-scale detectors, even down to the energy threshold. The predicted uncertainty remains unchanged when subselecting on those \TNGfifty{}~galaxies with merger histories similar to the Milky Way. Tabulated speed distributions, as well as Maxwell--Boltzmann fits, are provided for use in computing direct detection bounds or projecting sensitivities.

\vspace{\baselineskip}\noindent DOI: \href{https://doi.org/10.1103/wmpq-mw4h}{10.1103/wmpq-mw4h} \vspace{-\baselineskip}
\end{abstract}
\maketitle

\thispagestyle{fancy}
\section{Introduction}
\label{sec:1}
\nopagebreak 
Direct detection experiments seek to discover dark matter~(DM) through its interactions with a terrestrial detector (see, e.g., Refs.~\cite{Cerdeno10,Lin22,Yu25}). The scattering rate depends on both particle physics quantities, such as the interaction cross section, and astrophysical quantities, such as the DM speed distribution at the solar position. Uncertainties on the latter are assumed to be significant, but are rarely included in experimental results~\cite{McCabe10,Baxter21,Aalbers23a}, as they depend on the complicated evolution of the Milky Way~(MW) halo. The standard halo model~(SHM) simply assumes that DM near the Sun is equilibrated, following a Maxwell--Boltzmann speed distribution truncated at the Galactic escape speed~\citep{Wasserman86,Drukier86,Freese88}. Quantifying deviations from the SHM is necessary to robustly interpret experimental sensitivities. 

Cosmological simulations provide useful models for MW-like halos. Originally, DM-only simulations of such halos recovered speed distributions that underpredicted the mean DM speed relative to the SHM in the solar neighborhood~\cite{Wojtak05,Hansen06,Vogelsberger08,Vogelsberger09,Kuhlen10,Green10,Vogelsberger13,Bozorgnia17}. Later, hydrodynamical simulations showed that the addition of baryons deepens the gravitational potential of the galaxy, typically boosting DM speeds closer to the SHM prediction, but often discrepancies at the high-speed tail remain~\citep{Ling10,Tissera10,Pillepich14,Butsky16,Kelso16,Sloane16,Bozorgnia16,Bozorgnia17, Poole-McKenzie20, Bozorgnia20,Hryczuk20, Nunez-Castineyra23,Lawrence23,Staudt24}. Because of this predicted variability, constraints on the DM--nucleon interaction cross section are thought to have large uncertainty, especially for low-mass DM, where only high-speed material produces detectable signals. 

Simulated halos may have unreasonable DM speed distributions if their disk properties differ from those of our Galaxy, which is anomalously compact for its stellar mass~\citep{Hammer07,Licquia16a,Boardman20,Tsukui25}. A deeper potential would increase the circular speed of the disk and the DM speed near the Sun. Finding a simulated MW-like halo that closely reproduces our Galaxy's disk properties necessitates a large sample size, which is difficult given the computational expense of such simulations. Moreover, the rarity of finding a close match to the Galaxy makes it challenging to fully characterize the expected halo-to-halo variance for systems with different evolutionary histories.

This Letter uses \TNGfifty{}~\citep{Nelson19a,Pillepich19} to study a suite of 98 MW-like galaxies---the largest to date at this resolution. We implement a scaling of the simulated phase space, compressing the halos in position space, which correspondingly brings the circular speeds of each in line with observations, allowing for principled comparisons to the MW. Additionally, because the merger histories of the galaxies in this suite are known, we can test how the resulting speed distribution depends on the presence of a \emph{Gaia}~Sausage--Enceladus~(GSE) event~\cite{Helmi18,Belokurov18,Deason24} and the Large Magellanic Cloud~(LMC), two major events in the Galaxy's history. We show that, even when accounting for the halo-to-halo variance arising from different merger histories, the astrophysical uncertainty on direct detection limits is at or below the level of current experimental uncertainties for a characteristic ton-scale detector, e.g., Refs.~\cite{XENONCollaboration18,XENONCollaboration23,LZCollaboration25}.

\section{Simulated Milky Way--like galaxies}
\label{sec:2}
We use the highest-resolution simulation of the \Illustris\TNG{} suite, \TNGfifty{}~\citep{Nelson19a,Pillepich19}, which spans a volume of $(51.7~\text{Mpc})^3$. Halos are identified with \Subfind{}~\citep{Springel01,Dolag09} and given a total mass (\Mdyn{}) and a stellar mass (\Mstar{}). From this catalog, we select MW-like halos that (1)~have a stellar mass consistent with observations of the MW, viz. $(4$--$7.3)\times 10^{10}~\Msun$, (2)~are more than 500~kpc from any halo with larger \Mdyn{}, and (3)~are more than 1~Mpc from any halo with $\Mdyn > 10^{13}~\Msun$, as in Sec.~2.2 of Ref.~\citep{Folsom25}. The latter two requirements ensure the local environment of the MW-like halos is relatively empty, while still allowing for M31-like companions. We consider all particles near the MW-like halos, not only those which \Subfind{} assigns to the halo. The 98 halos satisfying these criteria are presented in Ref.~\citep{Folsom25}, along with a discussion of their merger histories.
 
\TNGfifty{} produces exponential disk lengths $R_d = 4.4_{-1.8}^{+3.2}$~kpc for MW-like systems~\citep{Sotillo-Ramos22,Pillepich24}, consistent with the observed $R_d\sim4$--8.5~kpc~\citep{Gadotti09,Pillepich24}. However, the MW is thought to have a more compact disk than other galaxies of its mass~\citep{Hammer07,Bovy13,Licquia16a,Boardman20,Tsukui25} (though see Ref.~\citep{Lian24}), with $R_d=1.7$--2.9~kpc~\citep{Hammer07,Juric08,Rix13,Bland-Hawthorn16,Pillepich24}, which is infrequently reproduced in \TNGfifty{}~\citep{Varma22,Rosas-Guevara22,Sotillo-Ramos22,Pillepich24}. 

\begin{figure}
 \centering
 \includegraphics{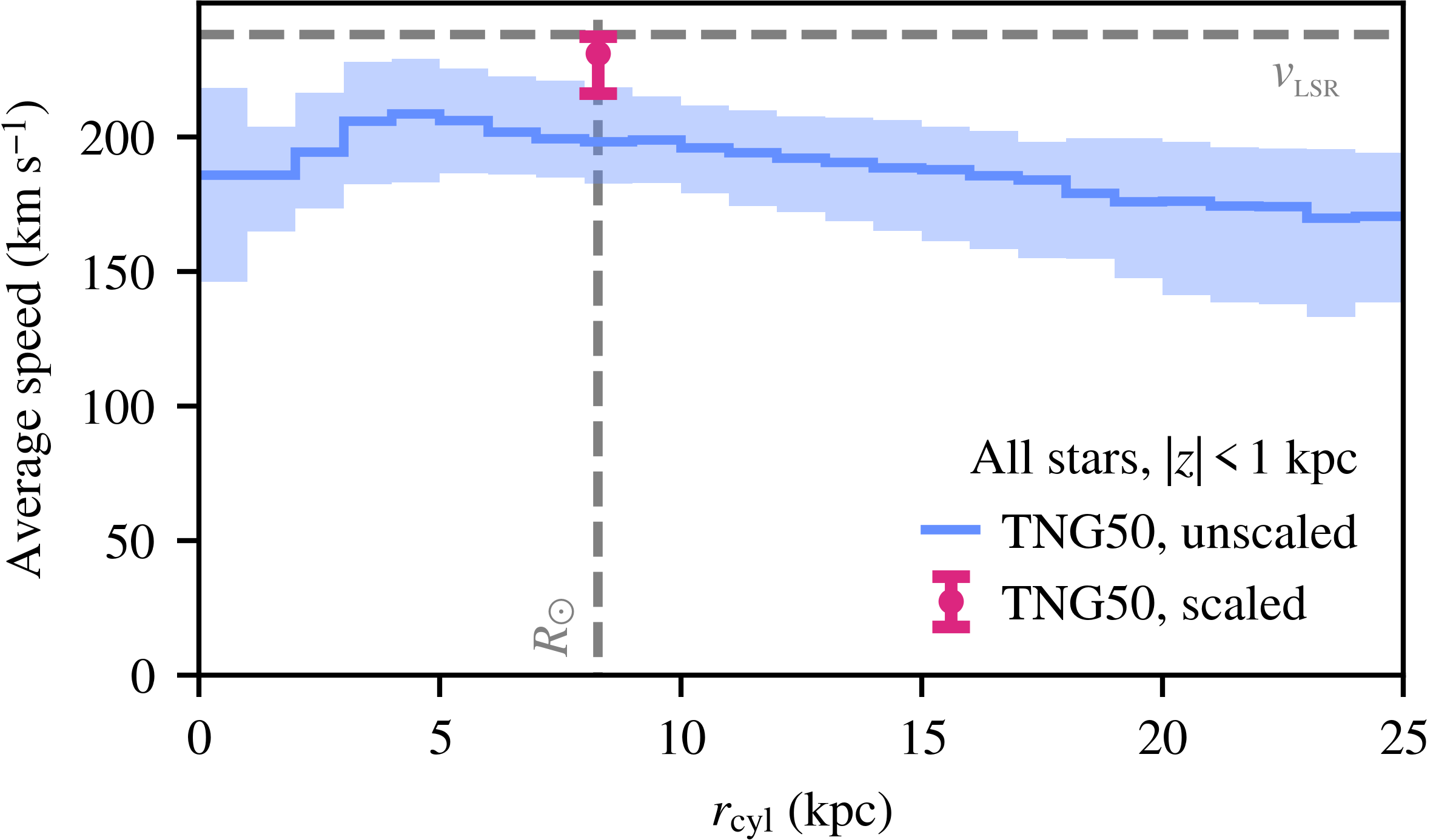}
 \caption{Average speed of stars in the disk plane (at heights $|z| < 1$~kpc) plotted as a function of cylindrical radius $r_\mathrm{cyl}$. The blue solid line indicates the median, evaluated binwise, of the \TNGfifty{} sample, while the shaded region indicates the 16th--84th percentile containment in each bin. The dashed gray horizontal and vertical lines correspond to the local standard-of-rest speed~($\vLSR = 238$~\kms) and the solar radius ($\Rsun = 8.3$~kpc), respectively. The Galaxy has a particularly compact stellar distribution. As such, \vLSR{} is 40~\kms{} faster than the median of the \TNGfifty{} galaxies at $\Rsun$. The scaling procedure developed in this Letter corrects for this, boosting stellar speeds closer to \vLSR{} at \Rsun{}. The pink vertical band brackets the 16th--84th percentile of the simulated galaxies after scaling, with a marker at the median.}
 \label{fig:1}
\end{figure}

Because the stellar mass distributions in the \TNGfifty{} sample are extended relative to the MW, stars in these galaxies are typically slower than the local standard-of-rest~(LSR) speed, $\vLSR = 238\pm1.5$~\kms{}~\citep{Reid04,Bland-Hawthorn16,GRAVITYCollaboration21,Baxter21}. \hyperref[fig:1]{Figure 1} shows the average speed of simulated stars near the disk plane ($|z| < 1$~kpc) as a function of cylindrical radius $r_\mathrm{cyl}$. The $z$~axis is parallel to the angular momentum of stars and star-forming gas within twice the stellar half-mass radius~\citep{Pillepich24}. The blue line indicates the median across the sample of MW-like galaxies, evaluated binwise, while the blue shaded region marks the 16th and 84th percentiles. At the solar radius, $\Rsun{} = 8.3$~kpc~\citep{GRAVITYCollaboration21,GRAVITYCollaboration24}, the average stellar speed in \TNGfifty{} is $198^{+20}_{-14}~\kms{}$ (16, 50, 84th percentile), 40~\kms{} slower than \vLSR{}. While some discrepancy is expected due to asymmetric drift~\citep{Binney08,Bovyprep.}, this effect is on the order of a few kilometers per second for most populations in the MW's disk~\citep{Schonrich10,Kawata19,Poder23,Ou25}, and the \TNGfifty{} stellar speeds are too low to be explained by this effect alone. 

The concerns outlined here apply to any simulation. Modeling subgrid baryonic physics is often a source of systematic uncertainty, especially for baryonic mass distributions, but any simulation that accurately reproduces observed spiral galaxy scaling relations will produce compact MW-like galaxies only as outliers, as the MW itself is not typical. 

\section{Scaling phase space}
\label{sec:3}
Because of the discrepancy between stellar speeds in the \TNGfifty{} sample and the observed solar speed, these systems will not make realistic predictions for the DM distribution in our Galaxy. For example, assuming that an observer is located at $\Rsun$ in an arbitrary \TNGfifty{} galaxy and inferring the DM speed distribution from the simulated particles at this radius will yield slower DM speeds relative to what should be expected for the MW. One na\"ive solution is to place the observer in the simulated halo at some radius $r \neq \Rsun$ in a way that scales with the size of the system (e.g., Ref.~\citep{Poole-McKenzie20}). However, this still rarely gives stellar speeds comparable to $\vLSR$, as can be seen from \autoref{fig:1}.

Previous works have approached this challenge differently. Reference~\citep{Lawrence23}, which studied one MW-like galaxy from the Latte suite~\citep{Wetzel16,Wetzel23}, introduced a boost to the observer's frame to match $\vLSR$. While the correction was small, this procedure corresponds to a transformation that does not conserve energy. Reference~\citep{Bozorgnia16} selected galaxies from the EAGLE~\citep{Schaye15,Crain15} and APOSTLE~\cite{Sawala16,Fattahi16} projects, requiring that they closely match the Galaxy's rotation curve. While principled, this selection reduced $\osim 60$ halos to 14. 

We aim to preserve the full sample of 98 MW-like systems in \TNGfifty{} while bringing them into alignment with the observed properties of our Galaxy in the solar neighborhood. Therefore, we perform an energy-conserving scaling to both position and velocity coordinates as follows:
\begin{enumerate}
\item With Newton's constant $G$, define the mass inside \Rsun{} as
 \begin{equation}
 \MLSR = \vLSR^2 \Rsun/G = 1.09\times 10^{11}~\Msun{}.
 \label{eq:1}
 \end{equation}
\item For each simulated halo, find the radius $\Rm$ that encloses a mass equal to \MLSR{}, and
\item scale phase space so $\Rm \mapsto \Rsun$, conserving energy,
\begin{equation}
 \vb{r} \mapsto  \frac{\Rsun}{\Rm}\;\vb{r}\quad\text{and}\quad \vb{v} \mapsto  \sqrt{\frac{\Rm}{\Rsun}}\;\vb{v}
\end{equation}
for simulated positions $\vb{r}$ and velocities $\vb{v}$.
\end{enumerate}
Any scaling of positions and velocities of this form is purely a change in units: it preserves the collisionless Boltzmann equation and therefore the evolution of the distribution function, as well as the expression for circular speed assumed in \autoref{eq:1} (see \hyperref[app:A]{Appendix A}). 

The particular choice of $\Rm\mapsto\Rsun$ ensures that the leading-order evolution of the distribution function is equivalent to that expected for the MW. Consider the gravitational potential $\phi(r)$ under a multipole expansion. After scaling, the monopole term evaluated at $\Rsun$ is $\phi(\Rsun) = G\MLSR{}/\Rsun$, as inferred for the MW. The higher-order terms of $\phi$ reflect the merger histories of individual halos and contribute to the variance we measure. 

\begin{figure}
 \centering
 \includegraphics{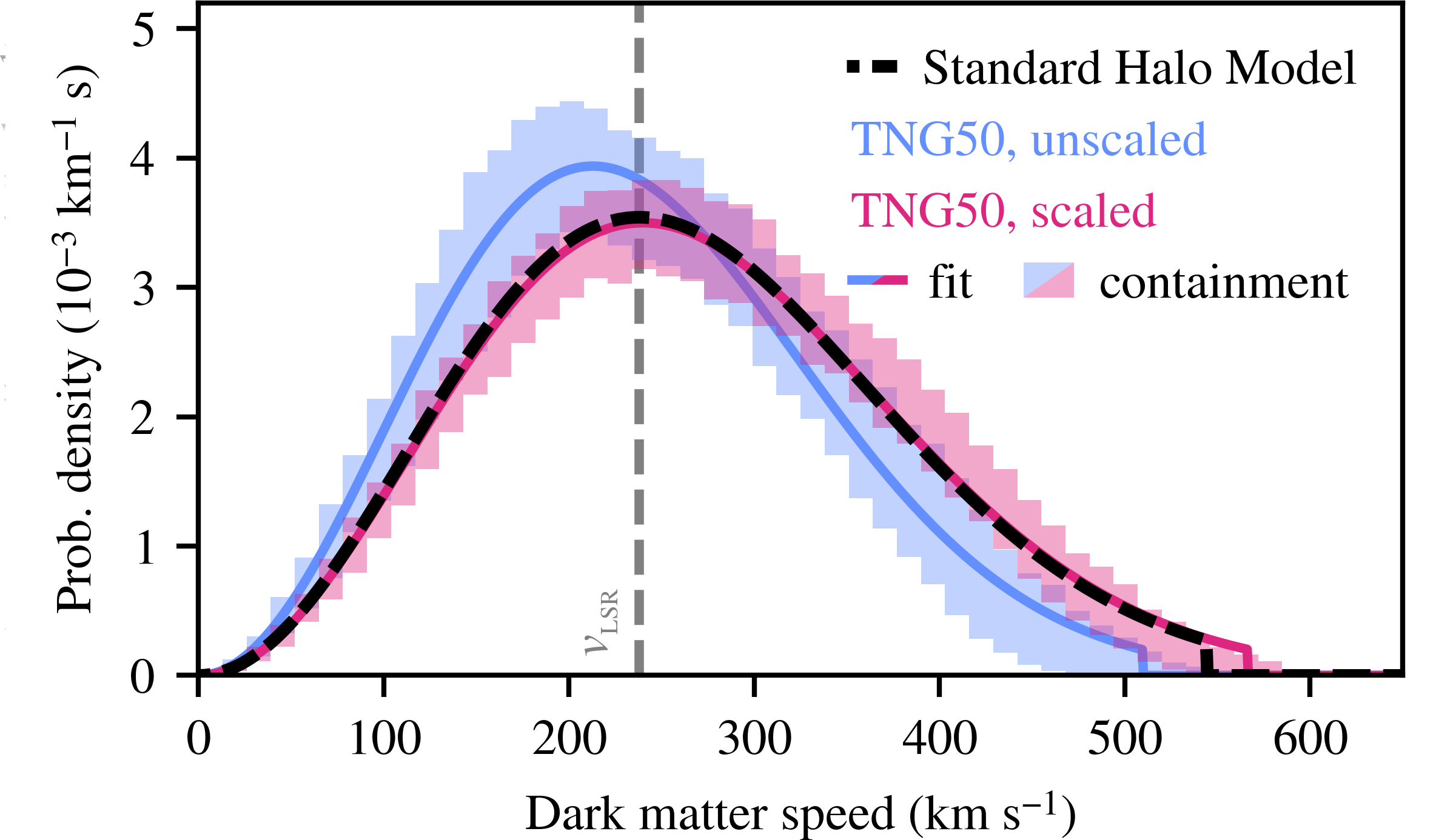}
 \caption{DM speed distributions around $\Rsun$, both before~(blue) and after~(pink) scaling the simulated halos. The shaded bands are binwise containment regions, bracketing the 16th--84th percentile of the distributions. The solid blue and pink lines show best-fit truncated Maxwell--Boltzmann probability distributions (see text) for the scaled and unscaled halos, respectively, and the black dashed line indicates the SHM prediction. The DM particles in the scaled halos have higher speeds, corresponding to their more compact mass distributions. With these higher speeds, the scaled halos better recover the SHM's peak at \vLSR, though their shapes are, in general, non-Maxwellian.}
 \label{fig:2}
\end{figure}

Because 92\% of halos in the \TNGfifty{} sample have enclosed masses at \Rsun{} below \MLSR{}, distances are typically shortened, with scale factor $\Rsun/\Rm = 0.72_{-0.08}^{+0.13}$. This compression increases the mass enclosed at \Rsun, yielding the correct $\phi$, and the boost in speeds corresponding to this new potential recovers \vLSR. The vertical pink line in \autoref{fig:1} shows the range of average stellar speeds at $\Rsun$ in the \TNGfifty{} galaxies after scaling; it is $231_{-15}^{+6}~\kms{}$, in better agreement with the MW. 

The DM speeds at $\Rsun$ exhibit a similar shift, shown in \autoref{fig:2}. For each halo, we construct the speed distribution from DM particles in the solar annulus, the cylindrical region where $|r_\mathrm{cyl}-\Rsun | < 1$~kpc and $|z|<1$~kpc. The shaded regions span the 16th--84th percentiles of the distributions, computed binwise, for the halos before~(blue) and after~(pink) scaling. The black dashed curve is the SHM (including a cutoff at the escape speed~544~\kms{}), and the vertical dashed gray line indicates the peak of the distribution at $v_0\equiv\vLSR = 238$~\kms{}. As containment regions, the shaded areas are not themselves probability distributions. To guide the eye, solid lines indicate truncated Maxwell--Boltzmann distributions fit to the \TNGfifty{} sample before and after scaling. The best-fit $v_0$ value is $213\pm 19$~\kms{} for the unscaled halos and $240\pm11$~\kms{} for the scaled halos, with a best-fit $\vesc{}$ value of $510 \pm 46$~\kms{} and $567\pm 43$~\kms{}, respectively (see \hyperref[app:B]{Appendix B}). Like for the stars, the unscaled DM speeds are too slow, with best-fit $v_0$ for the unscaled halos 25~\kms{} below \vLSR{}. After scaling, the speed distributions are in better agreement with the SHM, with a best-fit $v_0$ only 2~\kms{} away from \vLSR{}. This is expected, as the scaling is designed to recover a mean speed close to $\vLSR$. We note, however, that the individual halos' speed distributions are not necessarily Maxwellian: the distributions of the cylindrical velocity components have excess kurtosis $(-0.40^{+0.17}_{-0.14}, -0.11^{+0.20}_{-0.23}, -0.02^{+0.13}_{-0.15})$ for $(v_r,v_\phi,v_z)$.

We have validated this procedure by comparing the distribution of scaled halos to the unscaled halos with $\Rm\sim \Rsun$, i.e., those that are already as compact as the MW. For this subset, both the DM densities and the speed distributions in the solar annulus are consistent with those for the full sample of scaled galaxies, see \autoref{fig:A1}. Additionally, our prediction for the DM speed distribution is consistent with results from the burstier FIRE-2 model~\cite{Hopkins18}, but only once appropriate care is taken to match the observed circular speed (see \hyperref[app:A]{Appendix A}).

\section{Implications for direct detection}
\label{sec:4}

The changes to the phase-space distribution of DM around \Rsun{} induce changes in the elastic DM--nucleon recoil spectrum. We compute the recoil spectrum using the \textsc{wimprates} package~\citep{Aalbers23}, which we modify to accept distributions other than the SHM. The speed distributions are boosted to the reference frame of Earth on March 9, 2000, as per Refs.~\citep{McCabe14,OHare20,Baxter21}. The resulting recoil spectra are passed through an approximation to the XENON1T search pipeline provided in Ref.~\citep{XENONCollaboration22}, which emulates the analysis of 1~ton-yr of exposure from the XENON~Collaboration~\citep{XENONCollaboration18}, including background and detector response models. This pipeline constrains the DM--nucleon cross section \xsec{} that would exceed the modeled detector background at 90\% confidence limit~(C.L.). The search region in recoil energy---i.e., recoil energies that are reconstructed with at least 1\% efficiency---is from 2.25 to 58~keV.

\begin{figure}[t]
 \centering
 \includegraphics{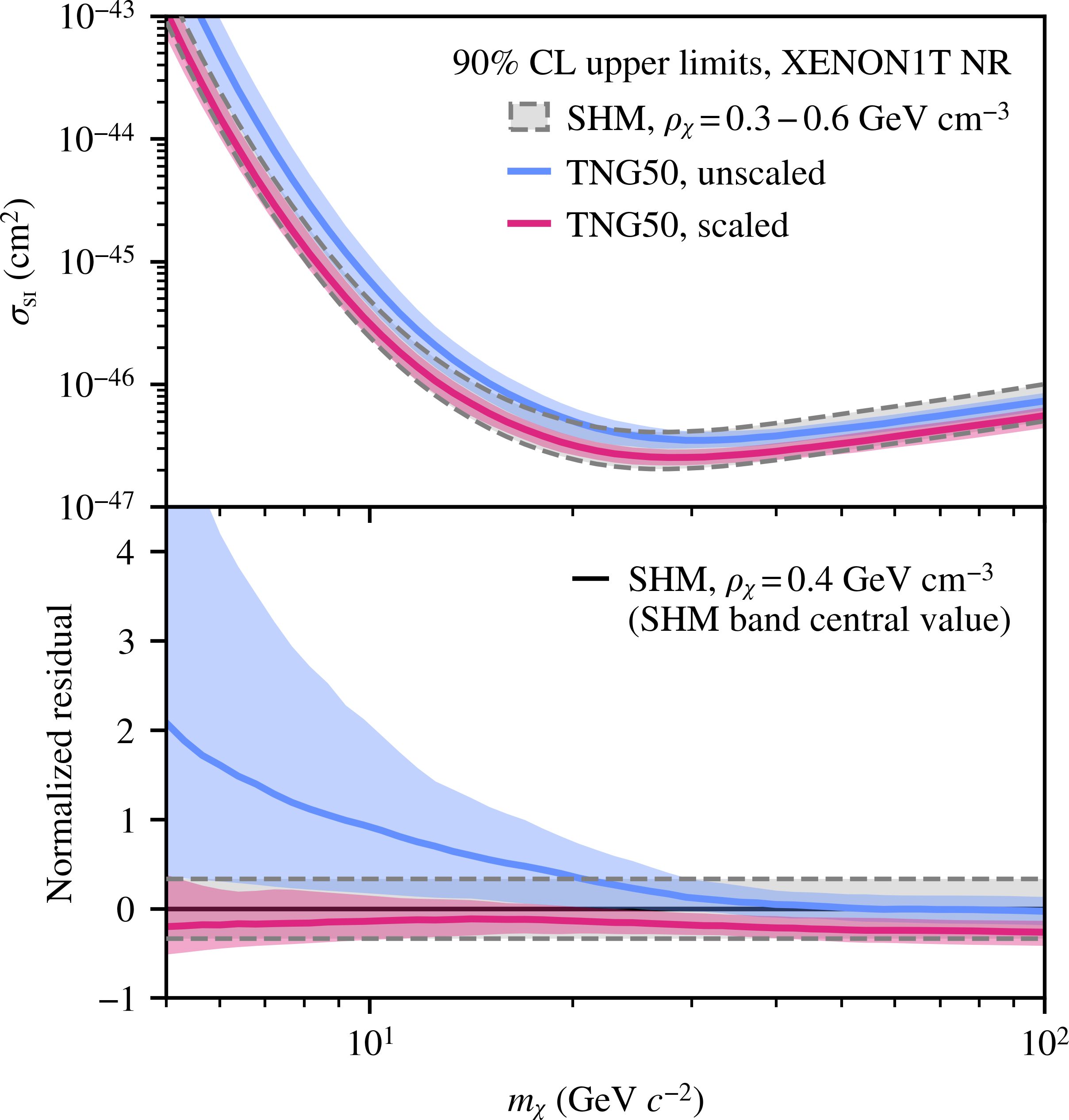}
 \caption{DM--nucleon spin independent cross section limits (90\% C.L.) as a function of DM mass $m_\chi$, as derived from simulated DM halos~(blue), compared to the predictions from the SHM (shown in gray, with the spread set by uncertainty in the local DM density $\rho_\chi$). The lower panel shows the fractional difference of the above curves relative to the SHM with a fiducial choice of $\rho_\chi = 0.4$~\GeVcm{}. The pink curves show limits derived from the same halos after phase-space scaling. While the overall normalization scales with $\rho_\chi$, variations in $\rho_\chi$ do not change the shape of the limit. The weakening of the unscaled limit at $m_\chi \lesssim 10$~\GeVc{} arises because the DM speed distribution is systematically slower than the SHM, an effect that scaling corrects for.}
 \label{fig:3}
\end{figure}

The top panel of \autoref{fig:3} shows the 90\% C.L. on the cross section for DM--nucleon scattering for the XENON1T experiment, as a function of the DM mass $m_\chi$. The shaded gray region indicates the bound set by the SHM, with a width set by observational uncertainty in the local DM density $\rho_\chi = 0.3$--0.6~\GeVcm{}~\cite{DeSalas21}. The blue band is the 16th--84th percentile containment region for the equivalent constraints set using the DM densities and speeds directly from \TNGfifty{}, and the pink band shows this after the aforementioned scaling. The lower panel shows the difference between these constraints and the prediction from the SHM with $\rho_\chi$ fixed to 0.4~\GeVcm{}, divided by the SHM prediction. The unscaled limits grow weaker than the SHM prediction at low DM masses and the uncertainty in the limit (i.e., the width of the shaded 1$\sigma$ inclusion region) grows large in this regime. In contrast, the limits set using the scaled halos show a smaller uncertainty and are in better agreement with the SHM. Specifically, the 1$\sigma$ containment region for $\xsec{}$ upper limits at $m_\chi = 5$~\GeVc{} ranges 56\% (123\%) below (above) the median for the unscaled phase-space distributions, shrinking to 39\% (73\%) in the scaled case.

The DM density sets the normalization of the sensitivity curves in \autoref{fig:3}, with higher $\rho_\chi$ leading to more stringent limits. The unscaled halos are less compact, have lower $\rho_\chi$, and set weaker limits than their scaled counterparts (see \hyperref[app:A]{Appendix A} for more detail on the DM densities). Beyond $m_\chi \sim 25$~\GeVc, most DM scattering events deposit enough energy to be detected, regardless of their speed. Therefore, uncertainty in $\xsec{}$ is primarily from $\rho_\chi$, which leads to a 33\% spread of the 1$\sigma$ containment region. Below this mass, the uncertainty is primarily from the speed distribution. As seen in \autoref{fig:2}, the unscaled speed distributions fall off much faster than the SHM at high speeds, while the scaled distributions still have support near the escape speed of the halo. The deficit of high-speed DM in the unscaled halos leads to a weakening of the \xsec{} limit, as there are few scattering events that deposit energies above the detector threshold. The scaled speed distributions, however, do not suffer this deficit, and the \xsec{} limit better matches that set by the SHM. 
\begin{figure}[b]
 \centering
 \includegraphics{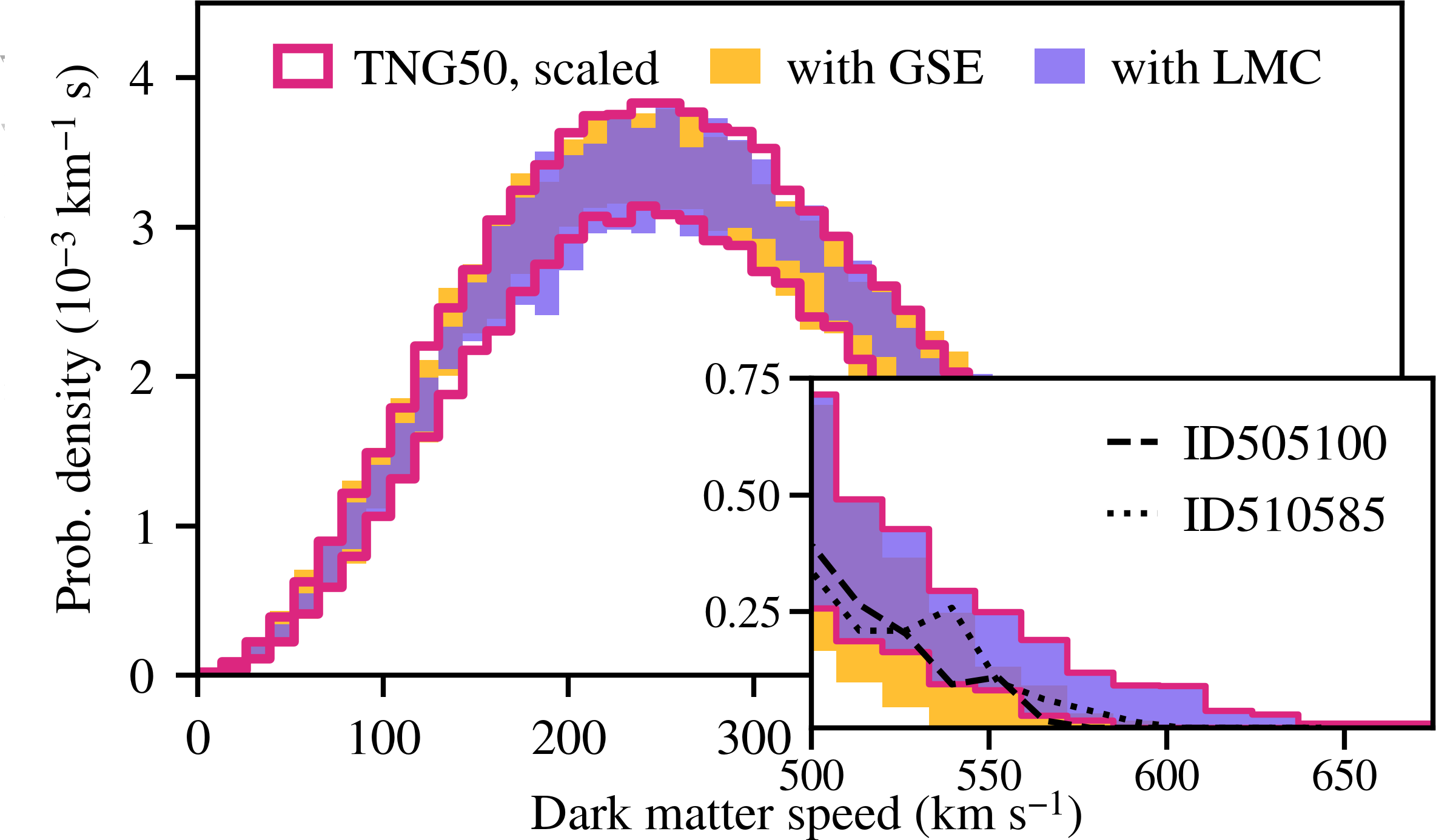}
 \caption{Scaled DM speed distributions for MW-like halos in the sample with GSEs~(yellow) and with LMCs~(purple). For comparison, the solid pink lines outline the shaded pink region from \autoref{fig:2} and bracket the range for all scaled halos. The bands correspond to the 16th--84th percentile containment. The inset focuses in on the high-speed tail, with the dashed and dotted black lines indicating the only two halos that experienced both a GSE and LMC event. A GSE~(LMC) event biases the high-speed tail to lower~(higher) values, although the shifts are relatively small and largely contained within the halo-to-halo uncertainty of the full sample.}
 \label{fig:4}
\end{figure}

We note that this analysis may not resolve small-scale DM streams or other local velocity structures because (1)~we construct the speed distribution as an azimuthal average of particles within an annulus around the galactic center, and (2)~the simulations are subject to a 288~pc softening length. Additionally, we neglect the gravitational effect of the Sun, which can focus the local DM phase space~\cite{Lee14}. These assumptions can potentially induce changes to the predicted scattering rate. 

\section{Restricting merger history}
\label{sec:5}
The large sample of halos allows us to determine the effects of the merger history on the speed distributions. We select one group of 32 halos that have had a GSE-like merger and another group of 11 halos with LMC-like satellites. Here, GSE-like mergers are identified as those that deposit stars with highly radially biased velocities that comprise 50\% of the \emph{ex situ} stars (defined as in Ref.~\citep{Folsom25}) within vertical distance $|z| \in [9, 15]$~kpc~\citep{Belokurov18,Deason18,Myeong18,Lancaster19,Necib19a,Fattahi19,Iorio21}. LMC-like satellites are those with mass $\Mdyn > 10^{11}~\Msun$ within the MW's virial radius (the $R_\mathrm{dyn}$ defined in Ref.~\cite{Folsom25}). \hyperref[fig:4]{Figure 4} shows the speed distribution for those galaxies with a GSE~(yellow band) and with an LMC~(purple band). Although the distributions for these two subsamples are consistent with the overall population (bracketed by the pink lines) to within 1$\sigma$, there are systematic shifts between them. In particular, the halos with an LMC-like satellite have slightly more support at the high-speed tail of the distribution, with $0.6^{+0.4}_{-0.4}\%$ of DM particles at or above the SHM escape speed, compared to $0.1^{+0.7}_{-0.1}\%$ in the overall population. This is consistent with results from Refs.~\citep{Besla19, Donaldson22, Smith-Orlik23}. In contrast, halos with a GSE-like merger yield slightly less support on the high-speed tail, with $0.2^{+0.3}_{-0.2}\%$ of DM particles at or above 544~\kms{}, though these halos are otherwise consistent with the overall population. This behavior differs from what was observed by Ref.~\cite{Bozorgnia20} using the Auriga simulations~\citep{Grand17}, although their sample consisted of only four halos with GSE-like events and thus may not have captured the full range of possibilities. In an upcoming study, we will compare the exact speed distributions from \TNGfifty{} with versions derived from GSE stellar debris (e.g., Refs.~\cite{Necib19a, Evans18,Zhu24}).

\section{Conclusions}
\label{sec:concl}
This Letter provides an updated and state-of-the-art theoretical prediction for the local speed distribution of the Milky Way, derived from the largest set of simulated galaxies to date at its resolution. We performed a physically motivated scaling of the positions and velocities of the particles in 98 MW-like halos in the \TNGfifty{} simulation to better reproduce the local circular speed \vLSR{} in our Galaxy. This scaling dramatically increased the number of MW-like halos that could be used to construct DM speed distributions and which could be robustly compared to the observationally motivated SHM. 
In general, the ensemble of recovered distributions exhibited excellent agreement with the SHM, even though the individual halos' distributions are not always well fit by a Maxwell--Boltzmann parametrization.
The corresponding 1$\sigma$ uncertainty from halo-to-halo variance is comparable to, and sometimes less than, the systematic experimental uncertainty for experiments such as LUX-ZEPLIN and XENON1T. Subselecting on those galaxies with merger histories similar to the Milky Way did not significantly affect the uncertainty range. These results motivate renewed efforts to model other sources of experimental uncertainty, such as the form factors and detector response functions near threshold, which had previously been thought to be subdominant. 

\hyperref[app:B]{Appendix B} presents tabulated distributions and analytic parametrizations for computing bounds or sensitivities for existing and future direct detection experiments. 
See \hyperref[app:B]{Appendix B} and Ref.~\cite{data_release} for our own data products, including the individual halos' DM speed distributions and densities, as well as a global fit to the DM speed distribution inferred for the MW. All \Illustris\TNG{} data are publicly available \footnote{\href{https://tng-project.org/}{https://tng-project.org/}}.

\vspace{\baselineskip}
\section{Acknowledgements}
We thank Akaxia Cruz, Jonah Rose, Sandip Roy, and Tal Shpigel for useful conversations. The work of C.B.~was supported in part by NASA through the NASA Hubble Fellowship Program Grant No. HST-HF2-51451.001-A awarded by the Space Telescope Science Institute, which is operated by the Association of Universities for Research in Astronomy, Inc., for NASA, under Contract No. NAS5-26555, as well as by the European Research Council under Grant No. 742104. M.L. and D.F. are supported by the Department of Energy~(DOE) under Award No. DE-SC0007968. M.L. is also supported by the Simons Investigator in Physics Award. D.F. is additionally supported by the Joseph H. Taylor Graduate Student Fellowship. L.N. is supported by the Sloan Fellowship, the NSF CAREER No. 2337864, and NSF Award No. 2307788. L.H. acknowledges support from the Simons Foundation through the ``Learning the Universe'' collaboration. The computations in this paper were run on the FASRC cluster supported by the FAS Division of Science Research Computing Group at Harvard University. The \Illustris\TNG{} simulations were undertaken with compute time awarded by the Gauss Centre for Supercomputing~(GCS) under GCS Large-Scale Projects GCS-ILLU and GCS-DWAR on the GCS share of the supercomputer Hazel Hen at the High Performance Computing Center Stuttgart~(HLRS), as well as on the machines of the Max Planck Computing and Data Facility~(MPCDF) in Garching, Germany.

\vspace{\baselineskip}
\section{Data availability}
The data that support the findings of this article are openly available \cite{data_release}.

\bibliography{main}

\titleformat{\section}[runin]{\itshape\addtocounter{section}{1}}{\thesection}{0pt}{Appendix \Alph{section}:\enspace}[---]
\setcounter{section}{0}

\onecolumngrid
\begin{center}
\large\textbf{End Matter}
\end{center}
\renewcommand{\theequation}{(\Alph{section}\arabic{equation})}
\twocolumngrid
\section{Validation of procedure}
\phantomsection
\label{app:A}

\setcounter{equation}{0}

\begin{figure*}
 \centering
\includegraphics{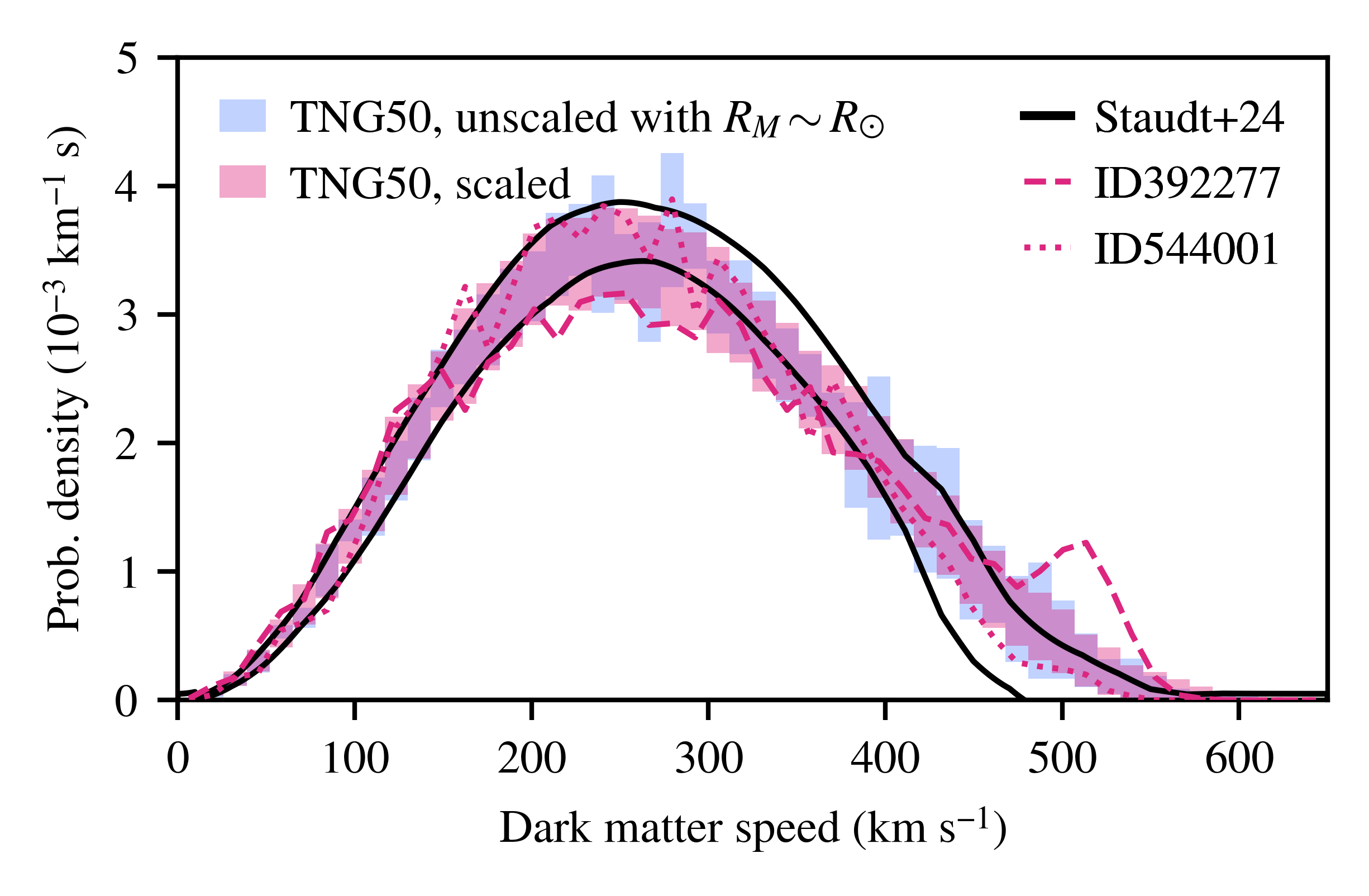} 
\includegraphics{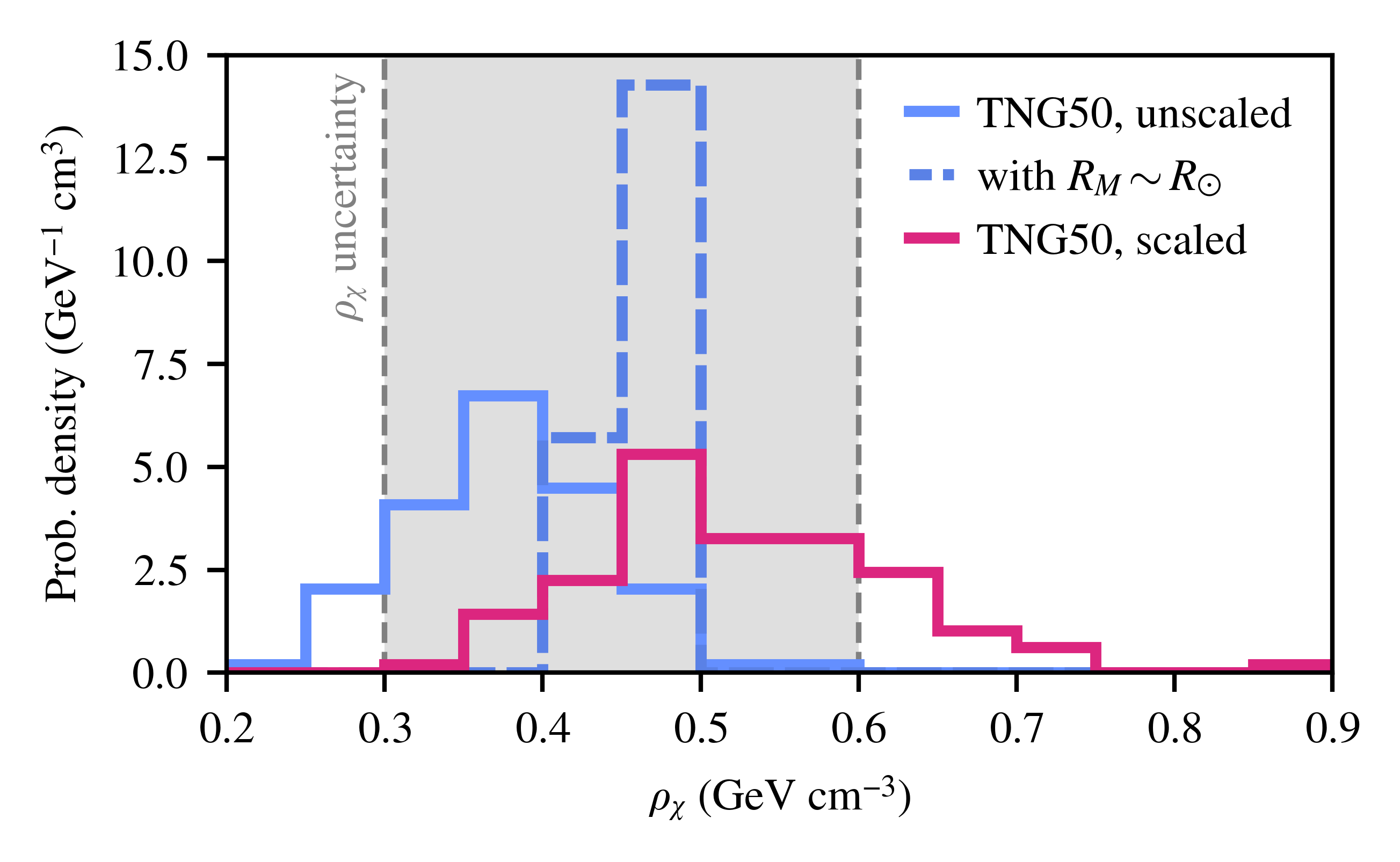}
 \caption{Probability distributions for DM speed (left) and density ($\rho_\chi$, right) around $\Rsun$. The left panel shows the pink scaled-halo band exactly as in \autoref{fig:2}, but the blue distribution representing the unscaled halos now only includes those with the smallest scaling factors, where $|\Rm{}/\Rsun - 1| < 0.125$. The right panel shows, in blue, the distribution of $\rho_\chi$ values for the unscaled halos, the full distribution with a solid line, and the least-scaled halos drawn with a dashed line. The pink solid line indicates the distribution for the scaled halos, and the shaded gray region indicates measurements of $\rho_\chi$ in the MW~\cite{DeSalas21}. In both panels, the resulting distribution of scaled halos is compatible with the most MW-like unscaled halos. The left panel also shows (in black) the $1\sigma$ band for the speed distribution inferred for the MW from Ref.~\citep{Staudt24}, which uses the FIRE-2 simulations. It agrees with our scaled distribution, suggesting that the conclusions of this Letter may be insensitive to the choice of subgrid model. 
 Additionally, we show scaled speed distributions for two individual halos that over- and underestimate the high-speed tail relative to the SHM in dashed and dotted pink, respectively, to emphasize that individual halos have non-Maxwellian speed distributions.
 }
 \label{fig:A1}
\end{figure*}

The evolution of a phase-space distribution function $f(\vb{x},\vb{v};\,t)$ is described by the collisionless Boltzmann equation, which takes the form
\begin{equation}
 \frac{\partial f}{\partial t} + \vb{v}\cdot\frac{\partial f}{\partial \vb{x}} - \frac{\partial\phi}{\partial \vb{x}} \cdot \frac{\partial f}{\partial \vb{v}} = 0
 \label{eq:A1}
\end{equation}
for particles evolving under a Hamiltonian $H = \frac{1}{2}v^2 + \phi(\vb{x},t)$. Consider a point $\vb{Q} = (\vb{x},\vb{v})$ under a coordinate transformation such that $\vb{Q}\mapsto \tvb{Q} = (\tvb{x}, \tvb{v}) = (\alpha^2\vb{x}, \alpha^{-1}\vb{v})$, with tildes indicating quantities evaluated in the new coordinate system. To ensure that $\tvb{v}$ is the time derivative of $\tvb{x}$, we must also take $\tilde{t} = \alpha^3t$.

By conservation of probability, $\tilde{f}(\tvb{Q}) = \alpha^{-3}f(\vb{Q})$, from the Jacobian of this coordinate transformation. Specifically, the nonzero entries of the Jacobian matrix are
\begin{equation}
 \frac{\partial{x_i}}{\partial{}\tilde{x}_j} = \delta_{ij}\alpha^{-2} \qquad\text{and}\qquad \frac{\partial{v_i}}{\partial{}\tilde{v}_j} = \delta_{ij} \alpha \, ,
\end{equation}
with $\delta_{ij}$ the Kronecker delta. From this, we can immediately determine how the first two terms transform:
\begin{equation}
 \frac{\partial \tilde{f}}{\partial \tilde{t}}\Big|_{\tvb{Q}} = \alpha^{-6}\frac{\partial f}{\partial t}\Big|_{\vb{Q}},\quad \tvb{v}\cdot\frac{\partial \tilde{f}}{\partial \tvb{x}}\Big|_{\tvb{Q}} = (\alpha^{-1}\vb{v})\cdot\Big(\alpha^{-5}\frac{\partial f}{\partial \vb{x}}\Big|_{\vb{Q}}\Big) \, .
\end{equation}

For the third term, let $\phi$ be the gravitational potential. The potential $\mathrm{d}\phi$ sourced by a mass $\mathrm{d}m$ at position $\vb{x}'$ is
\begin{equation}
 \mathrm{d}\phi(\vb{x}) = \frac{G\,\mathrm{d}m}{|\vb{x}' - \vb{x}|} \, .
\end{equation}
In scaled coordinates, with the mass at position $\tvb{x}'$, this becomes
\begin{equation}
 \mathrm{d}\tilde{\phi}(\tvb{x}) = \frac{G\,\mathrm{d}m}{|\tvb{x}' - \tvb{x}|} = \frac{G\,\mathrm{d}m}{\alpha^{2}|\vb{x}' - \vb{x}|} = \alpha^{-2}\mathrm{d}\phi(\vb{x}) \, .
\end{equation}
This holds for all masses, so 
\begin{equation}
 - \frac{\partial\tilde\phi}{\partial \tvb{x}} \cdot \frac{\partial \tilde f}{\partial \tvb{v}}\Big|_{\tvb{Q}} = - \Big(\alpha^{-4}\frac{\partial\phi}{\partial \vb{x}}\Big|_{\vb{Q}}\Big) \cdot \Big(\alpha^{-2}\frac{\partial f}{\partial \vb{v}}\Big|_{\vb{Q}}\Big) \, ,
\end{equation}
and each term in \autoref{eq:A1} transforms with a factor of $\alpha^{-6}$. 

As such, the form of the collisionless Boltzmann equation is unchanged in the new coordinate system: $\tilde{f}$ is out of equilibrium (i.e., $\tilde{f}$ has nonzero time derivative) if and only if $f$ is out of equilibrium, and an observer using the transformed coordinate system will observe gravitational evolution under $\tilde{\phi}$ proceeding as in the original coordinate system. Additionally, for a system with mass $M$ enclosed at a radius $R$, an observer in the transformed coordinate system will measure this mass to be enclosed within a radius $\tilde{R} = \alpha^2R$. The circular speed at this radius is $v(R) = \sqrt{GM/R}$, as in \autoref{eq:1}, and the observer will see 
\begin{equation}
 \tilde{v}(\tilde{R}) = \sqrt{\frac{GM}{\tilde{R}}} = \sqrt{\frac{GM}{\alpha^2 R}} = \alpha^{-1} v(R) \, ,
\end{equation}
consistent with the transformation assumed for $\tilde{v}$.

These arguments hold regardless of the value of $\alpha$, and one is free to choose any value for $\alpha$. The choice taken in this work, $\alpha = \sqrt{\Rsun/\Rm}$, ensures that the enclosed mass at $\tilde{x} = \Rsun$ is $\MLSR$ and therefore ensures that the leading-order term in $\tilde{\phi}$ is consistent from halo to halo. To verify that this scaling procedure still produces reasonable DM distribution functions, we compare the unscaled halos that are most MW-like, which undergo minimal scaling $(\alpha \sim 1)$, to the set of all 98 halos with scaling applied. The results of this test are shown in \autoref{fig:A1}. The left panel of \autoref{fig:A1} is formatted as \autoref{fig:2}, but with the unscaled distributions~(blue) restricted to halos with $\alpha \sim 1$. Specifically, the unscaled distributions show the seven halos with $|\Rm/\Rsun - 1| < 0.125$. Though the overall distributions of unscaled and scaled halos differ (as seen in \autoref{fig:2}), the scaled halos more closely match the most MW-like unscaled halos. 

Further, we reproduce the speed distribution from Ref.~\citep{Staudt24} (shown in black in \autoref{fig:A1}), who use 12 FIRE-2 galaxies~\cite{Hopkins18,Garrison-Kimmel19} to develop an inference pipeline for the local speed distribution of a MW-like galaxy, taking the local circular speed as an input. Their final prediction lies within the halo-to-halo variance of the \TNGfifty{} result. Although the statistics of the FIRE-2 sample are limited, this suggests that the conclusions of our Letter may not be highly sensitive to the details of the subgrid model.

The scaling procedure does modify $\rho_\chi$, the spatially averaged DM density within the solar annulus, as it brings particles toward the center of the halo. Prior to scaling, $\rho_\chi = 0.37_{-0.06}^{+0.05}$~\GeVcm{}; after scaling, this increases to $0.51_{-0.07}^{+0.13}$~\GeVcm{}. These values---both before and after scaling---are consistent with MW observations, which range from 0.3 to 0.6~\GeVcm{}~\cite{DeSalas21}. The full distributions are plotted in the right panel of \autoref{fig:A1}, where $\rho_\chi$ is shown before scaling~(blue) and after scaling~(pink). The dashed blue shows the subset of unscaled halos with $|\Rm/\Rsun - 1| < 0.125$, and again this distribution is in agreement with the scaled halos.

\begin{figure}
 \centering
 \includegraphics{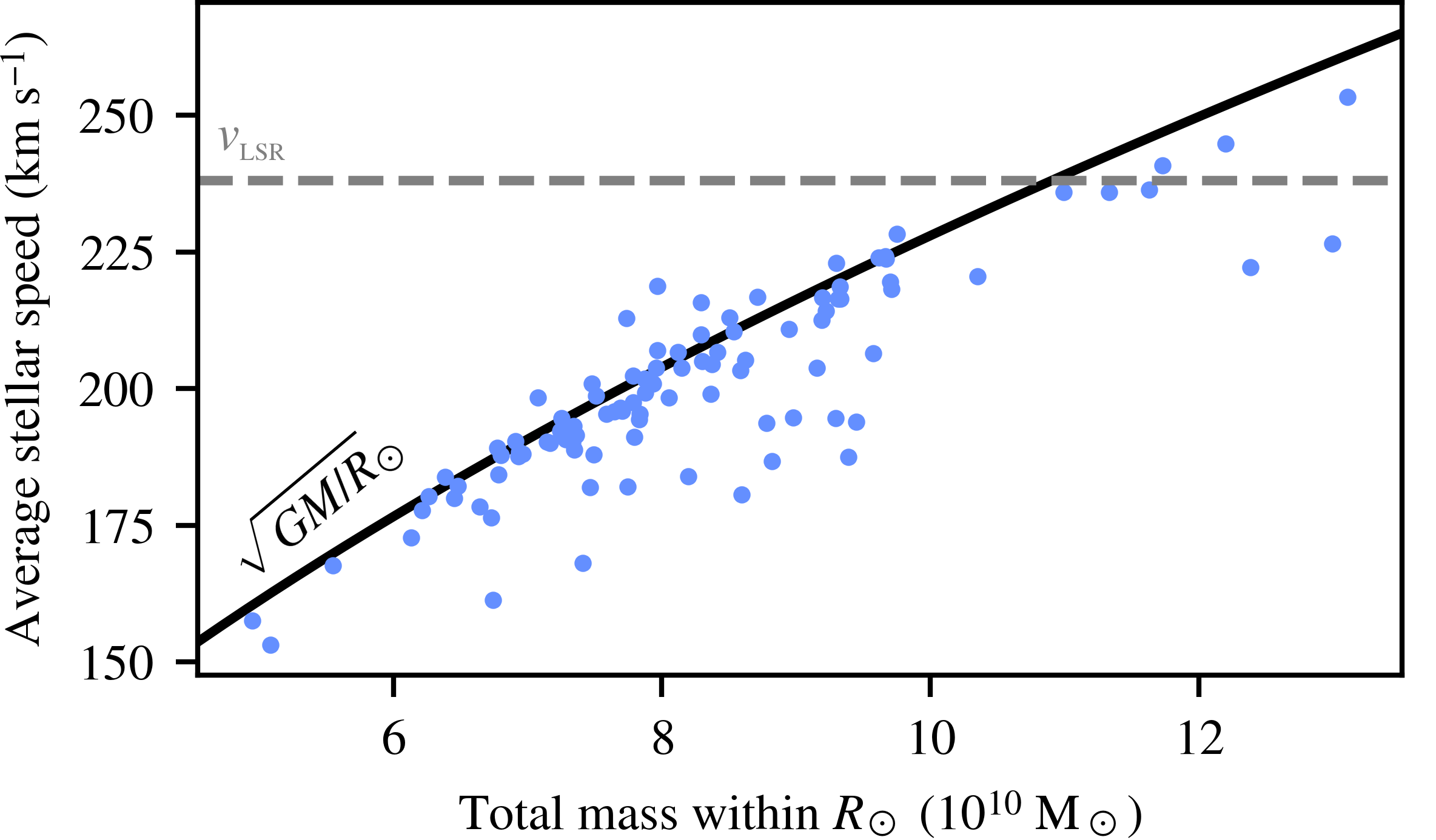}
 \caption{\label{fig:A2} For each MW-like galaxy, this shows the average speed of stars in an annulus around the solar radius (the same region used in \autoref{fig:2}), compared to the total mass $M$ enclosed within $\Rsun$. The speeds lag $\osim20$~\kms{} behind \vLSR{} (shown in dashed gray), and they correlate with the square root of the enclosed mass as expected, staying along the solid black curve which denotes $\sqrt{GM/\Rsun}$.}
\end{figure}

Finally, we also verify that the monopole term of $\phi$ sets the average speeds found in the simulation, and, therefore, that correcting this term should yield speeds comparable to $\vLSR$. This is shown in \autoref{fig:A2}, which plots the average speeds of stars in the solar annulus of the unscaled \TNGfifty{} halos against $M$, the mass enclosed within $\Rsun$. For reference, the local standard-of-rest speed is shown as a dashed gray horizontal line and a solid black line indicates $\sqrt{GM/\Rsun}$, the speed predicted by the monopole term of $\phi$. Average speeds follow this prediction to within $\osim3\%$ on average. These deviations may be caused by higher-order terms in the potential or out-of-equilibrium effects. Therefore, accounting for the discrepancy between the simulated masses within $\Rsun$ and the $\MLSR$ inferred for the MW should yield good models for the DM speed in the MW. 

\section{Recommendations for direct detection}
\phantomsection
\label{app:B}

\setcounter{equation}{0}
\setcounter{figure}{0}
\setcounter{table}{0}

Here, we provide expressions for the \TNGfifty{} DM distributions that can be used as an input to further studies, both on a halo-by-halo level and for the overall population of MW-like galaxies. 

Numerical interpolators for the exact speed distributions are available online~\cite{data_release}. For each halo, we provide the DM density around $\Rsun$. We also provide parametric speed distributions, in the form of a truncated Maxwell--Boltzmann,
\begin{gather}
 f_\mathrm{fit}(v) = 4\pi v^2 k\exp\left(-\frac{v^2}{v_0^2}\right) \Theta(\vesc{} - v),\qquad\text{with}\label{eq:B1}\\
 k^{-1} = (\pi v_0^2)^{3/2}\left[\operatorname{erf}\left(\frac{\vesc{}}{v_0}\right) - \frac{2}{\sqrt{\pi}} \frac{\vesc{}}{v_0}\exp\left(-\frac{\vesc{}^2}{v_0^2}\right)\right],\label{eq:b2}
\end{gather}
where $\Theta$ is the Heaviside step function, erf is the Gauss error function, $v_0$ is the peak speed, and $\vesc{}$ is the truncation speed.

For each simulated halo, we generate an ensemble of 10,000 speed distributions by randomly sampling (with replacement) DM speeds in the solar annulus. For each of these random samples, we perform a maximum likelihood fit to a truncated Maxwell--Boltzmann distribution to extract $v_0$ and $\vesc{}$. This bootstrapping technique quantifies the statistical uncertainty in fit parameters for the individual halos, which is typically on the order of 1~\kms{} in $v_0$ and 5~\kms{} in $\vesc{}$. These values are tabulated in the repository linked above, but we caution that the exact speed distributions are not always well modeled by this functional form. For example, the left panel of \autoref{fig:A1} shows an individual halo's speed distribution in dashed (dotted) pink that significantly over- (under-) predicts the high-speed tail with respect to its parametric fit.

An estimate for the globally preferred values of $v_0$ and $\vesc{}$ is determined by fitting a bivariate Gaussian distribution to the collection of all 98~000 sets of best-fit $v_0$ and $\vesc{}$. The result of this Gaussian fit is reported in \autoref{tab:B1}, where the means and standard deviations for $v_0$ and $\vesc{}$ are given in the first two columns of the table, and the third column is the Pearson correlation coefficient, i.e. the covariance normalized by the product of the individual standard deviations. The full covariance matrix can be reconstructed from these values. For readers interested in a single DM distribution to use for the MW, our recommendation is to use the values in this table, as they marginalize over the halo-to-halo variance present in the \TNGfifty{} sample.
\FloatBarrier
\begin{table}
 \setcounter{section}{2}
 \caption{\label{tab:B1}Best-fit values and uncertainty for the parameters of \autoref{eq:B1}. These are fit across the entire simulated sample, marginalizing over the halo-to-halo variance. For reference, recall that the SHM has $v_0 = 238~\kms{}$ and $\vesc{}=544~\kms{}$. The third column is the Pearson correlation coefficient, which (with the standard deviation on the individual parameters) determines the full covariance matrix.}
\begin{ruledtabular}
 \begin{tabular}{rccc}
 Sample & $v_0$ (\kms{}) & $\vesc{}$ (\kms{}) & Correlation \\
 \colrule
 TNG50, unscaled & $213\pm 19$ & $510 \pm 46$ & 0.75\\
 TNG50, scaled & $240\pm 11$ & $567 \pm 43$ & 0.60\\
 \end{tabular}
\end{ruledtabular}
\setcounter{section}{1}
\end{table}

\end{document}